\documentclass{article}
\usepackage{graphicx}
\usepackage[russian,english]{babel}

\evensidemargin=\oddsidemargin
\def\Title#1#2#3{%
    \baselineskip=18pt
    \begin{center}
          {\large\bf{#1} \\ }
          \bigskip\bigskip
          {#2} \\
          {#3} \\
    \end{center}}
\long\def\Abstract#1{%
         \bigskip
         \parbox{0.93\textwidth}{%
                 \begin{center}
                       {\bf Abstract} \\
                 \end{center}
                 \medskip{\baselineskip=14pt #1}
                 \vss}
         \bigskip}

\makeatletter
\renewcommand{\section}%
 {\@startsection{section}{1}{0pt}%
  {-3.25ex plus -1ex minus -.2ex}{1.5ex plus .2ex}%
  {\vspace*{5mm}\raggedright\large\bf }}
\renewcommand{\subsection}%
 {\@startsection{subsection}{2}{0pt}%
  {-2.25ex plus -.5ex minus -.2ex}{-1.5ex plus -.2ex}{\bf }}
\renewcommand{\subsubsection}%
 {\@startsection{subsubsection}{3}{0pt}%
  {-1.25ex plus -.2ex minus -.1ex}{-1.2ex plus -.2ex}{\bf }}

\makeatother

\begin{document}

\Title{The birth of the Universe as a result of the change\\
of the metric signature}%
{T. P. Shestakova}%
{Department of Theoretical and Computational Physics,
Southern Federal University,\\
Sorge St. 5, Rostov-on-Don 344090, Russia \\
E-mail: {\tt shestakova@sfedu.ru}}

\Abstract{In this paper, I discuss the idea that the birth of our Universe may be a result of a quantum transition from a physical continuum with the Euclidean signature to a Lorentzian spacetime. A similar idea was expressed by Andrei D. Sakharov. At the classical level, the idea was studied by George F. R. Ellis and his collaborators, who explored if solutions to the classical Einstein equation exist which admit a change of metric signature. The present paper aims at examining possible realizations of this idea at the level of quantum gravity, in the framework of the Wheeler -- DeWitt theory and in the extended phase space approach to quantization of gravity. I intend to answer the questions: Does the Hartle -- Hawking wave function imply such a realization? How can this idea be realized in the extended phase space approach to quantum gravity, where the change of signature is described by imposing special conditions on $g_{00}$-component of the metric in different regions of the physical continuum? The conclusion is that the idea can be realized from a formal mathematical point of view, but it can hardly help in understanding how spacetime structure and time itself appeared from a timeless continuum.}

\section{Introduction}
The title of this Special Issue, ``Light on Dark Words'', has reminded me the following lines from the poem ``The Night before Christmas'' by the great Russian philosopher Vladimir Solovyov \cite{Solovyov}:
\begin{otherlanguage}{russian}
\begin{verse}
Родился в мире свет, и свет отвергнут тьмою,\\
Но светит он во тьме, где грань добра и зла.\\
Не властью внешнею, а правдою самою\\
Князь века осужден и все его дела.
\end{verse}
\end{otherlanguage}
In English translation,
\begin{verse}
Was born the Light in the World, and it was spurned by the darkness,\\
But shines it in the gloom, where the edge of the goodness and evil is.\\
Not by an external power, but by the truth itself\\
The Dark Lord is condemned and so are his deeds.
\end{verse}
(Translated by the author of this paper.)

In my student years, I was very interested in Russian philosophy and, in particular, in the life and works of Vladimir Solovyov. I knew that he had died prematurely, in the age of 47, in Uzkoye, the Moscow estate of the princes Trubetskoy. In 1999, the International conference on Cosmoparticle Physics ``COSMION-99'' was held in Uzkoye, where there was a sanatorium of the Russian Academy of Sciences at that time. The conference was organized by Maxim Yurievich Khlopov. Uzkoye was the place where I have acquainted with him.

After that, I met Professor Khlopov many times at various conferences, many of which were organized by him. Last winter he invited me to participate in the workshop ``Developing A. D. Sakharov Legacy in Cosmoparticle Physics'' that was a part of the 1st Electronic Conference on the Universe. I had read some works by Sakharov on gravitation and cosmology when I had been a student, about the same time when I had got acquainted with Russian philosophy. The Khlopov's invitation made me return to the papers of Sakharov and look at them from the viewpoint of the extended phase space approach to quantization of gravity, which I have been developing for many years. It has resulted in my paper \cite{Shest1}, where I commented on Sakharov's work ``Cosmological transitions with changes in the signature of the metric'' \cite{Sakh1}.

Sakharov's paper about the changes in the signature of the metric was written at the time when the interest in quantum cosmology was increasing. It was inspired, as Sakharov himself mentioned, by works of Vilenkin \cite{Vil1}, Hartle and Hawking \cite{HH} and others. Sakharov suggested that the birth of our Universe may be thought as a result of a quantum transition from a physical continuum with the Euclidean signature $(+,+,+,+)$ to a Lorentzian spacetime with the signature $(-,+,+,+)$. However, Sakharov has not given a detailed description of this transition. As mentioned in \cite{Alt}, later his idea became popular. It was actively discussed in 1990s (see, for example, \cite{AB} and references therein). Now, there are different approaches to the change of metric signature, some are based on the classical description \cite{Zhang} while others rely on the quantum consideration \cite{BB}.

The idea of quantum transitions from a physical continuum with the Euclidean signature to a Lorentzian spacetime originates from the technique known as the Wick rotation and used in ordinary quantum field theory to regularize a path integral. Accordingly, one can define the path integral over field configurations in Minkowski spacetime or in Euclidean space, in the latter, one of coordinates being the so-called ``imaginary'' time as a result of the transformation $t\rightarrow -iy$. It can be applied to gravitational field; then the path integral is defined over all metrics with Euclidean signature. Hawking believed that quantum theory of gravity should be formulated using path integrals over metrics with Euclidean signature \cite{HH}. However, some researchers consider ``Euclidean quantum gravity'' as a specific approach which differs from other approaches to quantization of gravity. As a further generalization, one comes to the idea about a physical continuum which includes regions with different signatures of the metric and transitions between them. In principle, regions with different signature may exist inside of our Universe \cite{Sakh1,Shest1}. At the present level of quantum gravity, one can hardly give an explanation what caused transitions with changes in the metric signature. The notion about these transitions still remains to be completely hypothetical, and so much the more when one implies a transition that resulted in the creation of the Universe; it must take place beyond time, so the notion of causality is not applicable.

\begin{figure}
\centering
\includegraphics[width=10 cm]{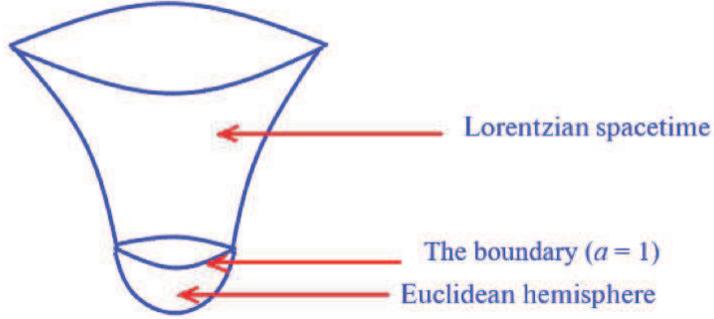}
\caption{\protect\small The physical continuum consisting of a Euclidean hemisphere glued to a Lorentzian spacetime on its boundary. The value of a scale factor $a=1$ at the boundary (in the Planck units). Cf. Figure 1 in \cite{Ellis}.\label{fig1}}
\end{figure}

In 1992, Ellis et al. published the paper \cite{Ellis} devoted to the change of the signature in the framework of general relativity. They called the idea, that the signature may change at very early times, one of the intriguing aspects of the Hartle -- Hawking ``no boundary'' proposal \cite{HH}. The result of the change would be the origin of the Universe from a region with the Euclidean signature, where there exists no time, and where there would be no boundary, according to the Hartle -- Hawking proposal. The region can be thought as a half of a Euclidean 4-sphere glued to a Lorentzian spacetime on its boundary, where the signature changes, as depicted at Figure~\ref{fig1}. For certainty, let us adopt that the change of signature takes place when a scale factor $a=l_{Pl}$ ($l_{Pl}$ is the Planck length); it is convenient to use the Planck units, then $a=1$. Ellis et al. investigated the question, if there are solutions of the Einstein equations that allow the change of the signature and describe the physical continuum outlined above. The authors considered such a solutions as a classical analog of what had been implied by the Hartle -- Hawking wave function of the Universe.

However, the ``no boundary'' proposal says that the wave function of the Universe should be given by a path integral over compact Euclidean geometries, so that the Universe does not have any boundaries, at least in the Euclidean regime. The Euclidean regime means a transition to imaginary time (the Wick rotation $t\rightarrow -iy$). Hawking admitted that the wave function can be analytically continued to the Lorentzian region \cite{Hawking1}, but does it imply a signature change? In his famous popular book \cite{Hawking2}, Hawking suggested that there are two descriptions of the history of the Universe: in real time and in imaginary time. He wrote that it is meaningless to ask, which time is real, ``real'' or ``imaginary'' time.

Though the authors of the paper \cite{Ellis} have not does not mentioned the paper by Sakharov \cite{Sakh1}, it seems that the consideration of a physical continuum consisting of regions with different signatures is closer to the ideas of Sakharov than to what was suggested by the ``no boundary'' proposal. Moreover, in \cite{Ellis}, the change of the signature is due to the change of the sign of $g_{00}$ component of the metric (which is denoted by $N$ in the paper by Ellis et al., but I prefer to keep this symbol for the lapse function). Since $g_{00}$ is supposed to be arbitrary, it can be fixed by some special condition (which, in fact, is a kind of a gauge condition for $g_{00}$). At the same time, the Hartle -- Hawking wave function satisfies the Wheeler -- DeWitt equation, the latter is known to be insensitive to the choice of $g_{00}$ (concerning the Wheeler -- DeWitt equation, see \cite{Rovelli}). Then, can one say that the Hartle -- Hawking wave function indeed describes the physical continuum consisting of the Euclidean and Lorentzian regions, while their boundary corresponding the change of the signature and the origin of the observed Universe?

Given the fact that the ``no boundary'' proposal was discussed mainly in the framework of the Wheeler -- DeWitt quantum geometrodynamics (see, for example, \cite{Vil2,HT1,Linde}), the proposal, as well as the signature change, can be considered from the viewpoint of other approaches to quantization of gravity. A good example is given in \cite{BB}, where the both are investigated based on loop quantum cosmology. In the present paper, I analyze the question from the viewpoint of the extended phase space approach to quantization of gravity \cite{SSV1,SSV2,SSV3,SSV4}. This approach crucially differs from the Wheeler -- DeWitt one. In this approach, it is argued that one does not have enough grounds to consider the latter gauge invariant, therefore, its main equation, the Wheeler -- DeWitt equation, loses its sense and should be replaced by the Schr\"odinger equation (this point is thoroughly discussed in \cite{Shest2}). The extended phase space approach is explicitly gauge noninvariant: The description of a gravitating system, be it a black hole or the whole Universe, is given from the viewpoint of a certain observer in a fixed reference frame. The metric signature in different regions of the physical continuum is determined by a special condition for $g_{00}$. This gives us an opportunity to discuss the signature change from a different perspective.

In the next sections, I consider the form of equations for the wave function of the Universe (the Wheeler -- DeWitt or Schr\"odinger equations) and their applicability in the Euclidean region. In Section 2, I discuss the generally accepted Wheeler -- DeWitt approach, and in Section 3, I analyse the question from the viewpoint of the extended phase space approach to quantization of gravity. The comparison of the two approaches and some preliminary conclusions will be given in Section 4.

Let us note that Ellis et al. have not discussed any possible reasons for the signature change. At the same time, a number of works have been published where hypotheses about the origin for the Lorentzian spacetime signature were put forward \cite{Greensite,CG,ORT}. Having accepted that the spacetime metric is subject to quantum fluctuations, one may suggest that the metric signature is subject to quantum fluctuations as well. The metric can be written as $g_{\mu\nu}=e_{\mu}^a\eta_{ab}e_{\nu}^b$, where $\eta_{ab}={\rm diag}(e^{i\theta},1,\ldots,1)$ and $\theta$ is considered as a dynamical field. However, we do not know what governs its dynamics. In the present paper we shall discuss the signature change that resulted into the appearance of our Universe. Therefore, dynamical changing of the field $\theta$ must take place ``before'' the birth of our Universe and ``before'' the beginning of time itself. On the other hand, one may be tempted to introduce some another ``time'', in which the field dynamics is developing. It can lead to logical inconsistencies. In any case, introducing the new field requires additional assumptions \cite{Greensite}, so that in the present paper we shall not go into this question.

\section{The Wheeler -- DeWitt equation and its solution}
There are two definitions of the wave function: as a solution to the corresponding equation (Schr\"odinger or Wheeler -- DeWitt equation) or via a path integral over some class of metrics. Obviously, these two definitions must lead to the same result. It is ensured by a special procedure of derivation of the equation from the path integral, originally proposed by Feynman \cite{Feynman}. For the Schr\"odinger equation, the procedure starts from the following formula (see Eq. (18) in \cite{Feynman}):
\begin{equation}
\label{Feyn}
\psi\left(x_{k+1},t+\varepsilon\right)=
 \frac1A\int\!\exp\left[\frac i{\hbar}S\left(x_{k+1},x_k\right)\right]\psi\left(x_k,t\right)dx_k,
\end{equation}
where $x_k$ stands for values of (generalised) coordinates at a moment $t_k=t$, $x_{k+1}$ stands for values of coordinates at a moment $t_{k+1}=t+\varepsilon$, $S\left(x_{k+1},x_k\right)$ is the classical action on the path between the points $x_k$ and $x_{k+1}$. The further procedure consists in the choice of approximation of the classical action and expansion of the both sides of (\ref{Feyn}) as series in terms of power of $\varepsilon$. Correspondingly, one obtains the Schr\"odinger equation in the first order in $\varepsilon$.
Strictly speaking, the procedure is not straightforwardly applicable to derivation of the Wheeler -- DeWitt equation, that is an operator form of the Hamiltonian gravitational constraint. One should also bear in mind that the wave function satisfying this equation does not depend on time. Also, it does not depend on the lapse function $N$, which is supposed to be a non-physical variable. Therefore, Hartle and Hawking \cite{HH} have written:
\begin{equation}
\label{H-WDW}
0=\int Dg D\phi \left[\frac{\delta S}{\delta N}\right] \exp\left(i S[g,\phi]\right),
\end{equation}
here $Dg D\phi$ denotes path integration over gravitational and scalar fields. Then, based on the classical relation,
$\displaystyle\frac{\delta S}{\delta N}$ is replaced by the Hamiltonian constraint. The classical expression of the constraint should be written in the operator form, with the replacements of all momenta by corresponding operators. At this point one faces the problem of ordering just like in canonical quantization. Once the choice of ordering is made, the resulting operator can be taken out for the integral sign, and one obtains the Wheeler -- DeWitt equation. Hawking emphasized \cite{Hawking1} that the Wheeler -- DeWitt equation is the same whether it is derived from a path integral over Euclidean or Lorentzian metrics.

A similar approach was used by Barvinsky and Ponomariov \cite{BP}. In the both cases, no exact definition of the path integral as a limit of finite-dimensional integral, or the so-called skeletonization, was given. It means that the derivation of the Wheeler -- DeWitt equation is based on canonical quantization with its problem of ordering more than on path integral methods. Halliwell \cite{Hall} used a particular choice of skeletonization for a path integral with the Batalin -- Fradkin -- Vilkovisky effective action. He argued that, depending on the integration range of the lapse function $N$, one obtains a temporal Schr\"odinger equation (if the range of $N$ is 0 to $\infty$) or the Wheeler -- DeWitt equation (if $N$ is integrated from $-\infty$ to $\infty$). It is not quite clear, however, why the second choice should be made.

Anyway, the Wheeler -- DeWitt equations contains no information on the sign of $g_{00}$, and so do its solutions. In \cite{HH,Hawking1}, a closed isotropic model with a scalar field was considered. In \cite{Hawking1}, Hawking discussed a scalar field $\phi$ with mass $m$. To estimate the path integral, he used the semiclassical approximation and classical Euclidean field equations. In a region of minisuperspace in which $a^2m^2\phi^2<1$ ($a$ is a scale factor), the wave function grows exponentially. However, if $a>(m\phi)^{-1}$, there exist no real solution to the Euclidean field equations. In this situation, Hawking proposed to analytically extend the obtained approximate expression into the Lorentzian region. The resulting wave function grows exponentially in the Euclidean region ($a<(m\phi)^{-1}$) and oscillates in the Lorentzian region ($a>(m\phi)^{-1}$).

Let me note that it is just an interpretation suggested by Hawking that the wave function would give a probability amplitude for some Euclidean geometry in the certain region of minisuperspace (for small values of $a$) and for a Lorentzian spacetime in the other region (for large values of $a$). The line $a=(m\phi)^{-1}$ serves as a surface of the signature change, in spite of that the $g_{00}$ component of the metric is thought to be a redundant variable, and its sign has not been explicitly considered. The justification is that the Euclidean version of the Einstein equations is used on one side of the line $a=(m\phi)^{-1}$, while their Lorentzian form is used on the other side. One can accept this interpretation or not. Moreover, it seems to be model-dependent.

Since Ellis et al. studied solutions to classical field equations, they imposed special conditions on the $g_{00}$ component. They discussed two kinds of such conditions:
\begin{itemize}
\item A continuous change of signature, the example is given by $g_{00}(t)=t$.
\item A discontinuous change of $g_{00}$, for example, $g_{00}=-1$ in the Lorentzian region and $g_{00}=1$ in the Euclidean region.
\end{itemize}
In the next section we inquire if similar conditions could be incorporated into some version of quantum geometrodynamics.

\section{The extended phase space approach to quantization of gravity\\
and the signature change}
In \cite{Shest1}, I argued that the signature in different regions of the physical continuum can be fixed by special gauge conditions on components of the metric tensor. To take it into account in quantum cosmology, one should have a gauge-dependent equation for the wave function of the Universe. It seems to be in contradiction with the generally accepted requirement of gauge invariance of the theory. But can one satisfy this requirement in the same way as in quantum theory of non-gravitational fields?

It is well-known that a path integral over gauge fields is divergent. At first, authors of works on quantum cosmology did not pay much attention to this point. Later, they started to use the technique developed for gauge field theories. In these theories, to make the path integral converge, the original gauge-invariant action is replaced by some effective action which has the following structure:
\begin{equation}
\label{S_gauge}
S_{(eff)}=S_{(grav)}+S_{(gf)}+S_{(ghost)},
\end{equation}
Obviously, the gauge-fixing term $S_{(gf)}$ and the ghost term $S_{(ghost)}$ are not gauge-invariant, but gauge invariance can be restored by imposing asymptotic boundary conditions on Lagrange multipliers of gauge conditions and ghosts. At the physical level, the asymptotic boundary conditions correspond to asymptotic states, which are usually presumed in laboratory experiments, and play an important role in such approaches to quantization of gauge theories as the Faddeev -- Popov one \cite{FP} or the Batalin -- Fradkin -- Vilkovisky approach \cite{BFV1,BFV2,BFV3}. However, they are not justified for a gravitating system which does not posses asymptotic states (the only exception is the case of asymptotically flat spacetime). For a system without asymptotic states one cannot ensure gauge invariance of the theory. In this situation, the Wheeler -- DeWitt equation loses its sense. One cannot insist that it must be true equation for the wave function of the Universe \cite{Shest2}.

If we reject imposing the asymptotic boundary conditions, we should deal with a system, in which all degrees of freedom, including gauge and ghost ones, are treated on an equal footing. All them are involved into {\it extended} phase and configurational spaces. One can apply to the system the standard procedure of derivation of the Schr\"odinger equation from the path integral with the effective action (\ref{S_gauge}) \cite{SSV1,SSV2,SSV3,SSV4}. The result of this procedure will be an equation for the wave function of the Universe depending on gauge conditions. At the first stage we obtain the temporal Schr\"odinger equation
\begin{equation}
\label{SE1}
i\,\frac{\partial\Psi(N,q,\theta,\bar\theta;\,t)}{\partial t}
 =H\Psi(N,\,q,\,\theta,\,\bar\theta;\,t),
\end{equation}
where $H$ is the Hamiltonian operator in extended phase space,
\begin{equation}
\label{H}
H=-\frac1N\frac{\partial}{\partial\theta}
   \frac{\partial}{\partial\bar\theta}
  -\frac1{2M}\frac{\partial}{\partial Q^{\alpha}}MG^{\alpha\beta}
   \frac{\partial}{\partial Q^{\beta}}+U(N, q)-V[f];
\end{equation}
Here $q=\{q^a\}=\{a,\phi\}$ stands for physical variables, $\theta$, $\bar\theta$ are the Faddeev -- Popov ghosts, $N$ is the only gauge degree of freedom in the model (it may be, for example, the lapse function) that is subject to the condition $N=f(q)+k$, $k$ is some constant; $V[f]$ is a quantum correction to the potential $U(N,q)$, $M$ is the measure in the path integral;
\begin{equation}
\label{Galphabeta}
G^{\alpha\beta}=\left(
\begin{array}{cc}
f_{,a}f^{,a}&f^{,a}\\
f^{,a}&g^{ab}
\end{array}
\right);\quad
f_{,a}=\frac{\partial f}{\partial q^a};\quad
Q^{\alpha}=(N,\,q^a);
\end{equation}
$g^{ab}$ is an inverse metric of configurational subspace of physical variables.

Let us note that in the extended phase space approach, the Hamiltonian constraint, as well as all Einstein equations, are modified, since they are obtained from the effective action (\ref{S_gauge}) taking into account two additional terms, $S_{(gf)}$ and $S_{(ghost)}$. The modified Hamiltonian constraint can be written as $H=E$, where $H$ is the Hamiltonian in extended phase space and $E$ is a conserved quantity. It means that now the spectrum of the Hamiltonian operator is not limited by the only zero eigenvalue, in contrast with the Wheeler -- DeWitt approach. In its turn, it implies that one can avoid the static picture of the world, which is a well-known problem of the Wheeler -- DeWitt quantum geometrodynamics. In the extended phase space approach, one has a number of stationary quantum states, and, in principle, a quantum state of the Universe can change, for example, through quantum transitions. Therefore, without contradictions, one can consider the temporal Schr\"odinger equation as a candidate for a basic equation in quantum gravity.

The true origin of the operator $\displaystyle\frac{\partial}{\partial t}$ in (\ref{SE1}) is fixing a reference frame. At the mathematical level, it is done by introducing a gauge fixing term into the effective action. It leads to altering the gravitational constraints; not only the Hamiltonian constraint, but also the momentum constraints are modified. The gauge conditions are imposed on the lapse and shift functions and so determine the spacetime structure. It means that geometry of the Universe is described from the viewpoint of some observer in a certain reference frame including a chronometer.

A wave function -- a solution to the Schr\"odinger equation -- depends on physical, gauge and ghost variables. For our minisuperspace model its structure looks like
\begin{equation}
\label{GS}
\Psi(N,q,\theta,\bar\theta;t)
 =\int\Psi_k(q,t)\,\delta(N-f(q)-k)\,(\bar\theta+i\theta)\,dk.
\end{equation}
The wave function (\ref{GS}) contains information about geometry of the model as well as about the gauge condition which characterizes the state of the observer in accordance with the spirit of general relativity.

The function $\Psi_k(q,\,t)$ is of special interest for us. It describes a state of the physical subsystem for a reference frame fixed by the condition $N=f(q)+k$. The function is a solution to the equation
\begin{equation}
\label{phys.SE}
i\,\frac{\partial\Psi_k(q;\,t)}{\partial t}
 =H_{(phys)}[f]\Psi_k(q;\,t),
\end{equation}
which can be obtained by substituting (\ref{GS}) into (\ref{SE1}),
\begin{equation}
\label{phys.H}
H_{(phys)}[f]=\left.\left(-\frac1{2M}\frac{\partial}{\partial q^a}M g^{ab}\frac{\partial}{\partial q^b}
 +U(N, q)-V[f]\right)\right|_{N=f(q)+k}.
\end{equation}
In what follows, we shall ignore a small quantum correction $V[f]$, which is of $\hbar^2$ order. Now we have all necessary mathematics to describe the signature change in the framework of the extended phase space approach to quantization of gravity.

In \cite{Ellis}, Ellis et al. consider the Friedmann -- Lemaitre -- Robertson -- Walker (FLRW) spacetime and a scalar field with a potential $V(\phi)$. For simplicity, they put $\dot\phi=0$ (the dot denotes time derivative), that makes this model be equivalent to the FLRW universe with a cosmological constant. The spacetime interval for this model is
\begin{equation}
\label{FLRW}
ds^2=-N^2(t)dt^2+a^2(t)\left[d\chi^2+\sin^2\chi\left(d\theta^2+\sin^2\theta d\varphi^2\right)\right],
\end{equation}
and the action reads
\begin{equation}
\label{action-1}
S=\!\int\!dt\,\left[-\frac12\frac{a\dot a^2}N+\frac12Na
  -Na^3\Lambda+\pi\left(\dot N-\frac{d f}{d a}\dot a\right)
  +N\dot{\bar\theta}\dot\theta\right].
\end{equation}

The Schr\"odinger equation for the physical part of the wave function will look as (cf. \cite{Shest3}, Eq.(18)):
\begin{equation}
\label{Schr-eq-gen}
\left.\left[-\frac12\sqrt{\frac N a}\frac{\partial}{\partial a}
  \left(\sqrt{\frac N a}\frac{\partial\psi}{\partial a}\right)
 +\frac12 N a\psi-\Lambda Na^3\psi\right]\right|_{N=f(a)}
 =i\frac{\partial\psi}{\partial t}.
\end{equation}

\subsection{Discontinuous signature change}
At first, we shall turn to a discontinuous signature change, when $g_{00}=-1$ in the Lorentzian region and $g_{00}=1$ in the Euclidean region. According to (\ref{FLRW}), we can choose $N=1$ in the Lorentzian region (the value of a scale factor $a>1$ in the Plank units) and $N=i$ in the Euclidean region ($a<1$). For $N=1$, Eq. (\ref{Schr-eq-gen}) will be reduced to
\begin{equation}
\label{Schr-eq1}
-\frac 1{2a}\frac{\partial ^2\psi}{\partial a^2}
 +\frac 1{4a^2}\frac{\partial\psi}{\partial a}
 +\frac12 a\psi-\Lambda a^3\psi
 =i\frac{\partial\psi}{\partial t}.
\end{equation}
Then, one can go to the stationary Schr\"odinger equation as in ordinary quantum mechanics,
\begin{equation}
\label{Schr-eq1-st}
-\frac 1{2a}\frac{d^2\psi}{da^2}
 +\frac 1{4a^2}\frac{d\psi}{da}
 +\frac12 a\psi-\Lambda a^3\psi=E\psi.
\end{equation}

As I argued above, one cannot ensure that the Wheeler -- DeWitt equation is a true equation for the wave function of the Universe. It means that a spectrum of the Hamiltonian is not limited by the only zero eigenvalue. One can admit that the Universe can be created in a state with a non-zero eigenvalue. The assumption about non-zero energy of the Very Early Universe does not contradict the notion of quantum theory that some violation of the energy conservation law may occur as it takes place when particle-antiparticle pairs are created from a vacuum. However, one can expect that eventually the Universe would appear to be in a state with zero total energy. It may be a result of a series of quantum transitions between states with different energy levels, or it may be a result of a decay of the original state in which the Universe happened to be immediately after its nucleation. So, $E=0$ is an important particular case.

The equation (\ref{Schr-eq1-st}) has no singular point in the Lorentzian region. Before writing an analog of (\ref{Schr-eq1}) for the Euclidean region, one should ask the question: Does the Schr\"odinger equation make sense in a physical continuum with Euclidean signature? One should remember that, in an analog of (\ref{Feyn}) $t$ is not a temporal coordinate, but just one of coordinates of the physical continuum, say, $y$. One can distinguish this coordinate among others by that it can be continued into the Lorentzian region after the transformation $y\rightarrow it$. From a purely formal viewpoint, an analog of the left-hand side of (\ref{Feyn}) can be expanded to the first order in $\varepsilon$:
\begin{equation}
\label{lhs-Feyn}
\psi\left(x,y+\varepsilon\right)
 =\psi(x,y)+\varepsilon\frac{\partial\psi}{\partial y}.
\end{equation}
As for the right-hand side of (\ref{Feyn}), one should replace the action with an Euclidean one and take the so-called Gaussian quadratures, i.e. integrals like
\begin{equation}
\label{Gauss}
\int\limits_{-\infty}^{\infty}\xi^n\exp(i\alpha\xi^2)d\xi,\quad n=0,1,\ldots
\end{equation}
(cf. Eq.(26) in \cite{Feynman}). For a particle of mass $m$ moving in one dimension under a potential $V(x)$, one can adopt the simple approximation of the action
\begin{equation}
\label{approx}
S\left(x_{k+1},x_k\right)=\frac{m\varepsilon}2\left(\frac{x_{k+1}-x_k}{\varepsilon}\right)^2-\varepsilon V\left(x_{k+1}\right),
\end{equation}
so that $\xi=x_{k+1}-x_k$, $\alpha=\displaystyle\frac m{2\hbar\varepsilon}$. The integrals (\ref{Gauss}) need regularization. As a rule, one implies the following regularization:
\begin{equation}
\label{Gauss-reg}
\int\limits_{-\infty}^{\infty}\xi^n\exp(i\alpha\xi^2)d\xi=\lim_{\eta\to 0}\int\limits_{-\infty}^{\infty}\xi^n\exp\left(i(\alpha+i\eta)\xi^2\right)d\xi.
\end{equation}
Another way to regularize the integrals (\ref{Gauss}) is the transition $t\rightarrow -iy$ to the Euclidean action. For particles and non-gravitational fields, it makes the path integral be convergent and, correspondingly, the integrals (\ref{Gauss}) be well-defined. In the above example of the particle of mass $m$ one has
\begin{equation}
\label{Euclid-approx}
\frac i{\hbar}S\left(x_{k+1},x_k\right)\to -S_E\left(x_{k+1},x_k\right)
 =-\frac{m\varepsilon}2\left(\frac{x_{k+1}-x_k}{\varepsilon}\right)^2-\varepsilon V\left(x_{k+1}\right),
\end{equation}
However, the gravitational part of the Euclidean action is not positive-definite \cite{Hawking3}. In our minisuperspace model, it is manifested in the minus sign before the ``kinetic'' part
$\left(-\displaystyle\frac12\frac{a\dot a^2}N\right)$ in (\ref{action-1}) and, therefore, the integrals (\ref{Gauss}) are divergent. The Euclidean action for our model reads
\begin{equation}
\label{action-E}
-S_E=\!\int\!dt\,\left[\frac12\frac{a\dot a^2}N+\frac12Na
  -Na^3\Lambda+i\pi\left(\dot N-\frac{d f}{d a}\dot a\right)
  -N\dot{\bar\theta}\dot\theta\right].
\end{equation}

The problem of indefiniteness of the gravitational action has been discussed by many authors. In particular, Linde \cite{Linde} suggested to use for gravity the Wick rotation with the opposite sign $t\rightarrow iy$ instead of $t\rightarrow -iy$. This suggestion was criticized by Hawking and Turok \cite{HT2}. They argued that the action for gravitational fluctuations is positively-definite, in contrast with the gravitational background action, and there is no invariant way to treat the two separately.

In this situation, one has the following possibilities:
\begin{itemize}
\item One can conclude that, from a strictly mathematical point of view, the Schr\"odinger equation is not valid in the Euclidean region.
\item One can ignore the problem with the Wick rotation as a mathematical subtlety which is not essential and may be resolved by a future theory.
\item One can argue that there are no fluctuations and matter fields in the Euclidean region, so that one can use the Wick rotation with the opposed sign as a trick to ensure convergence of the integrals.
\end{itemize}

Let us choose the third possibility. The Feynman procedure needs a further generalization along the line mapped out in \cite{Cheng,SSV4}. The calculations lead to the equation
\begin{equation}
\label{Schr-eq-Euclid}
\left.\left[\frac i2\sqrt{\frac N a}\frac{\partial}{\partial a}
  \left(\sqrt{\frac N a}\frac{\partial\psi}{\partial a}\right)
 -\frac i2 N a\psi+i\Lambda Na^3\psi\right]\right|_{N=f(a)}
 =i\frac{\partial\psi}{\partial y}.
\end{equation}
Choosing the condition $N=i$ we come to the Eq. (\ref{Schr-eq1}) that we had in the Lorentzian region. Therefore, the wave function satisfies the same equation on both sides of the surface of signature change. The equation has the singular point $a=0$ in the Euclidean region. The wave function can be expanded into series in its neighbourhood:
\begin{equation}
\label{wf1}
\psi(a)=C_1\left(1+O\left(a^4\right)\right)+C_2a^{\frac32}\left(1+O\left(a^2\right)\right),
\end{equation}
where $C_1$ and $C_2$ are constants. Therefore, the wave function behaves like a regular function in the neighbourhood of this point.

\subsection{Continuous signature change}
Now I turn to the continuous change of signature. In \cite{Ellis}, the condition $g_{00}(t)=t$ was considered, however, in the extended phase space approach, it would be more convenient to use a condition of the form $N=f(a)$. So, we can take $N=\sqrt{a^2-l_{Pl}^2}$, or, in the Planck units, $N=\sqrt{a^2-1}$. Under this condition Eq.~(\ref{Schr-eq-gen}) reads
\begin{equation}
\label{Schr-eq2}
-\frac12\frac{\sqrt{a^2-1}}a\frac{\partial ^2\psi}{\partial a^2}
 -\frac 14\frac 1{a^2\sqrt{a^2-1}}\frac{\partial\psi}{\partial a}
 +\frac12 a\sqrt{a^2-1}\psi-\Lambda a^3\sqrt{a^2-1}\psi
 =i\frac{\partial\psi}{\partial t},
\end{equation}
or, going to the stationary Schr\"odinger equation,
\begin{equation}
\label{Schr-eq2-st}
-\frac12\frac{d^2\psi}{da^2}
 -\frac 14\frac 1{a(a^2-1)}\frac{d\psi}{da}
 +\left(\frac12 a^2-\Lambda a^4-E\frac a{\sqrt{a^2-1}}\right)\psi=0.
\end{equation}
This equation has two singular points: $a=0$ and $a=1$. In the case when $E=0$ one can expand its solutions in series. The expansion of the wave function in the neighbourhood of $a=0$ is given by (\ref{wf1}) again, and the expansion in the neighbourhood of $a=1$ is
\begin{equation}
\label{wf2}
\psi(a)=C_1\left[1+O\left((a-1)^2\right)\right]+C_2(a-1)^{\frac34}\left[1+O(a-1)\right].
\end{equation}
We can conclude that, in the both discontinuous and continuous cases, the wave function is regular at the point $a=0$ which corresponds to the classical singularity of the Einstein equations in the Lorentzian region, as well as at the point of the signature change $a=1$.

\section{Discussion}

In this paper, we explore the idea that our Universe appeared from a region of some abstract physical continuum where time does not exist. This idea implies the notion of the signature change, when one of the principal values of the metric tensor changes its sign. More specifically, we suppose that the sign of $g_{00}$ component of the metric tensor changes.

In the Wheeler -- DeWitt quantum geometrodynamics, the $g_{00}$ component has no physical significance; neither the Wheeler -- DeWitt equation nor its solutions depend on it. Therefore, in this theory, the signature change can be only implicitly implemented. We have seen the example of such implementation in the Hartle -- Hawking approach, when the path integral, that determines the wave function, was evaluated with solutions to the classical Einstein equations. Depending on the existence of real solutions, in one region of minisuperspace, the solutions to the Euclidean Einstein equations were used, while in the other region the ``Lorentzian'' solutions were required. Thus, in this approach, the sign of $g_{00}$ component is taken into account indirectly since it is important in the appropriate solutions. The boundary between two regions in minisuperspace plays the role of the surface of the signature change.

The creators of quantum geometrodynamics, Wheeler \cite{Wheeler} and DeWitt \cite{DeWitt}, emphasized that only 3-geometry makes sense. The wave function gives a probability amplitude for the Universe to have some 3-geometry, and it does not depend on 4-geometry in which this 3-geometry is embedded and which is fixed by values of the lapse and shift functions. In the Wheeler -- DeWitt approach, the lapse and shift functions, being gauge (non-physical) degrees of freedom, are believed to be irrelevant. But it means {\it the destruction of spacetime} in quantum geometrodynamics (see a more detailed discussion in \cite{Shest4}). The wave function of the Universe is declared to be {\it not depending} on the lapse and shift functions, which fix a reference frame. On the contrary, in general relativity, the theory we try to quantize, any solution of the Einstein equations can be obtained only {\it after fixing} some reference frame.

Therefore, in the Wheeler -- DeWitt approach, it does not matter if the 3-geometry is embedded in some spacetime or a 4-dimensional space with the Euclidean signature. But it must be significant for us as the observers since speaking about the birth of the Universe, we imply the appearance of spacetime. By definition, spacetime is a connected 4-dimensional manifold with a Lorentzian metric \cite{HE}. Accordingly, the space we observe cannot be embedded in a region of the physical continuum with the Euclidean signature, and the definition of the wave function as the path integral over Euclidean metrics should be considered as a mathematical trick, while the description of history of the Universe in imagery time being just a ``reflection'' of its history in real time (on accordance with the interpretation of Hawking \cite{Hawking2}).

In the extended phase space approach, the condition for $g_{00}$ can be taken into account explicitly. The form of the equation for the wave function depends on this condition. In the both cases of discontinuous and continuous signature changes, the point $a=0$ which corresponds to the classical singularity, is a singular point of the equations. In the continuous case, we have the second singular point $a=1$ which indicates the surface of the signature change. It may be essential that the both singular points correspond to the points of special physical interest in classical solutions to the Einstein equations, namely, the cosmological singularity and the signature change. Nevertheless, according to the preliminary analysis, the solutions are regular in all these points.

So, we have various possibilities to describe the signature change. However, any formal description is getting us no closer to understanding what happens at the point of the signature change at the physical level. A widely accepted point of view is that time does not exist in the realm of quantum gravity, and time emerges from a ``timeless universe'' at the semiclassical level. However, this approach contains a logical error, since it relies on the semiclassical approximation, but making use of the semiclassical approximation, in its turn, requires the existing of time (see \cite{CC} in this connection).

From the viewpoint of Russian philosophy, attempting to describe the beginning of the Universe, we try to understand the Unknowable. Semen Frank, the Russian philosopher, who developed the ideas of Vladimir Solovyov, wrote in his book ``The Unknowable'' \cite{Frank} that the problem of the origin of our World is incomprehensible in logical terms. We can construct some models similar to the mentioned above model of dynamical signature change \cite{Greensite,CG,ORT}, but these models create new questions. In this endless process of cognition, we can only grasp intuitively some aspects of Being.

The signature change is an event that takes place beyond time. One of the possibilities, recently discussed in the literature, is that it can be described as a spontaneous breaking of the signature symmetry \cite{ZMY}. At this point physical properties of space is radically restructured. It is a transition from a space with the Euclidean signature, where no physical effect can propagate, to a spacetime with its causal structure, that gives room to physical interactions including electromagnetic one. So, ``was born the Light in the World''\ldots


\begin{thebibliography}{99}
\bibitem{Solovyov}
\begin{otherlanguage}{russian}
V. S. Solovyov,
 ``Ночь на Рождество'' [``The Night before Christmas''],
 in: В. С. Соловьев, {\it Смысл любви. Избранные произведения}, ``Современник''. Москва, 1991, стр. 445
 [V. S. Solovyov, {\it The meaning of Love. Selected works}, ``Sovremennik'' Publishing House, Moscow, Russia, 1991, p. 445 (in Russian)].
\end{otherlanguage}
\bibitem{Shest1}
T. P. Shestakova,
 ``On A. D. Sakharov's hypothesis of cosmological transitions with changes in the signature of the metric'',
 {\it Universe} {\bf 7} (2021) 151.
\bibitem{Sakh1}
A. D. Sakharov,
 ``Cosmological transitions with changes in the signature of the metric'',
 {\it Zh. Eksp. Teor. Fiz.} {\bf 87} (1984) 375-383.
\bibitem{Vil1}
A. Vilenkin,
 ``Birth of inflationary universes'',
 {\it Phys. Rev.} {\bf D27} (1983) 2848-2855.
\bibitem{HH}
J. B. Hartle and S. W. Hawking,
 ``Wave function of the Universe'',
 {\it Phys. Rev.} {\bf D28} (1983) 2960-2975.
\bibitem{Alt}
B. L. Altshuler,
 ``Andrei Sakharov's research work and modern physics'',
 {\it Physics -- Uspekhi} {\bf 64} (2021) 427-451.
\bibitem{AB}
B. L. Altshuler and A. O. Barvinsky,
 ``Quantum cosmology and physics of transitions with a change of the spacetime signature'',
 {\it Physics -- Uspekhi} {\bf 39} (1996) 429-460.
\bibitem{Zhang}
F. Zhang,
 ``Alternative route towards the change of metric signature'',
 {\it Phys. Rev.} {\bf D100} (2019) 064043.
\bibitem{BB}
M. Bojowald and S. Brahma,
 ``Loop quantum gravity, signature change, and the no-boundary proposal''
 {\it Phys. Rev.} {\bf D102} (2020) 106023.
\bibitem{Ellis}
G. Ellis, A. Sumeruk, D. Coule and C. Hellaby,
 ``Change of signature in classical relativity'',
 {\it Class. Quantum Grav.} {\bf 9} (1992) 1535-1554.
\bibitem{Hawking1}
S. W. Hawking,
 ``The quantum state of the Universe'',
 {\it Nucl. Phys.} {\bf B239} (1984) 257-276.
\bibitem{Hawking2}
S. W. Hawking,
 {\it A Brief History of time.}
 Bantam Books, London, UK, 1988.
\bibitem{Rovelli}
C. Rovelli,
 ``The strange equation of quantum gravity'',
 {\it Class. Quantum Grav.} {\bf 32} (2015) 124005.
\bibitem{Vil2}
A. Vilenkin,
 ``Quantum cosmology and the initial state of the Universe'',
 {\it Phys. Rev.} {\bf D37} (1988) 888-897.
\bibitem{HT1}
S. W. Hawking and N. Turok,
 ``Open inflation without false vacua'',
 {\it Phys. Lett.} {\bf B425} (1998) 25-32.
\bibitem{Linde}
A. D. Linde,
 ``Quantum creation of an open inflationary universe'',
 {\it Phys. Rev.} {\bf D58} (1998) 083514.
\bibitem{SSV1}
V. A. Savchenko, T. P. Shestakova and G. M. Vereshkov,
 ``Quantum geometrodynamics of the Bianchi IX model in extended phase space'',
 {\it Int. J. Mod. Phys.} {\bf A14} (1999) 4473-4490.
\bibitem{SSV2}
V. A. Savchenko, T. P. Shestakova and G. M. Vereshkov,
 ``The exact cosmological solution to the dynamical equations for the Bianchi IX model'',
 {\it Int. J. Mod. Phys.} {\bf A15} (2000) 3207-3220.
\bibitem{SSV3}
V. A. Savchenko, T. P. Shestakova and G. M. Vereshkov,
 ``Quantum geometrodynamics in extended phase space -- I. Physical problems of interpretation and mathematical problems of gauge invariance'',
 {\it Gravitation \& Cosmology} {\bf 7} (2001) 18-28.
\bibitem{SSV4}
V. A. Savchenko, T. P. Shestakova and G. M. Vereshkov,
 ``Quantum geometrodynamics in extended phase space -- II. The Bianchi IX model'',
 {\it Gravitation \& Cosmology} {\bf 7} (2001) 102-116.
\bibitem{Shest2}
T. P. Shestakova,
 ``Is the Wheeler - DeWitt equation more fundamental than the Schrödinger equation?''.
 {\it Int. J. Mod. Phys.} {\bf D27} (2018) 1841004.
\bibitem{Greensite}
J. Greensite,
 ``Dynamical origin of the Lorentzian signature of spacetime'',
 {\it Phys. Lett.} {\bf B300} (1993) 34-37.
\bibitem{CG}
A. Carlini and J. Greensite,
 ``Why is spacetime Lorentzian?'',
 {\it Phys. Rev.} {\bf D49} (1994) 866-878.
\bibitem{ORT}
S. D. Odintsov, A. Romeo and R. W. Tucker,
 ``Dynamical generation of spacetime signature by massive quantum fields on a topologically non-trivial backgrouud'',
 {\it Class. Quantum Grav.} {\bf 11} (1994) 2951-2959.
\bibitem{Feynman}
R. P. Feynman,
 ``Space-time approach to non-relativistic quantum mechanics'',
 {\it Rev. Mod. Phys.} {\bf 20} (1948) 367--387.
\bibitem{BP}
A. O. Barvinsky and V. N. Ponomariov,
 ``Quantum geometrodynamics: The path integral and the initial value problem for the wave function of the Universe'',
 {\it Phys. Lett.} {\bf B167} (1986) 289--294.
\bibitem{Hall}
J. J. Halliwell,
 ``Derivation of the Wheeler -- DeWitt equation from a path integral for minisuperspace models'',
 {\it Phys. Rev.} {\bf D38} (1988) 2468--2481.
\bibitem{FP}
L. D. Faddeev and V. N. Popov,
 ``Feynman diagrams for the Yang -- Mills field'',
 {\it Phys. Lett.} {\bf B25} (1967) 29-30.
\bibitem{BFV1}
E. S. Fradkin and G. A. Vilkovisky,
 ``Quantization of relativistic systems with constraints''.
 {\it Phys. Lett.} {\bf B55} (1975) 224-226.
\bibitem{BFV2}
I. A. Batalin and G. A. Vilkovisky,
 ``Relativistic S-matrix of dynamical systems with boson and fermion constraints'',
 {\it Phys. Lett.} {\bf B69} (1977) 309-312.
\bibitem{BFV3}
E. S. Fradkin and T. E. Fradkina,
 ``Quantization of relativistic systems with boson and fermion first- and second-class constraints'',
 {\it Phys. Lett.} {\bf B72} (1978) 343-348.
\bibitem{Shest3}
T. P. Shestakova,
 ``On the meaning of the wave function of the Universe'',
 {\it Int. J. Mod. Phys.} {\bf D28} (2019) 1941009.
\bibitem{Hawking3}
S. W. Hawking,
 ``The path-integral approach to quantum gravity'',
 in: {\it General relativity. An Einstein centenary survey}, eds. by S. W. Hawking and W. Israel, Cambridge University Press, Cambridge, UK, 1979, pp. 746--789.
\bibitem{HT2}
S. W. Hawking and N. Turok,
 ``Comment on `Quantum Creation of an Open Universe', by Andrei Linde'',
 arXiv: gr-qc/9802062.
\bibitem{Cheng}
K. S. Cheng,
 ``Quantization of a general dynamical system by Feynman's path integration formulation'',
 {\it J. Math. Phys.} {\bf 13} (1972) 1723--1726.
\bibitem{Wheeler}
J. A. Wheeler,
 {\it Einsteins vision},
 Springer Verlag, Berlin -- Heidelberg -- New York, 1968.
\bibitem{DeWitt}
B. S. DeWitt,
 ``Quantum theory of gravity. I. The canonical theory'',
 {\it Phys. Rev.} {\bf 160} (1967) 1113--1148.
\bibitem{Shest4}
T. P. Shestakova,
 ``Is the Copenhagen interpretation inapplicable to quantum cosmology?'',
 {\it Universe} {\bf 6} (2020) 128.
\bibitem{HE}
S. W. Hawking and G. F. R. Ellis,
 {\it The large scale structure of space-time},
 Cambidge University Press, Cambridge, UK, 1973.
\bibitem{CC}
E. Y. S. Chua and C. Callender,
 ``No Time for Time from No-Time'',
 {\it Philosophy of Science} {\bf 88} (2021) 1172-1184.
\bibitem{Frank}
S. L. Frank,
 {\it The Unknowable: An ontological introduction to the philosophy of religion},
 English translation by B. Jakim,
 Ohio University Press, Athens, USA, 1983.
\bibitem{ZMY}
A. R. Ziyaee, M. Mohsenzadeh and E. Yusofi,
 ``A possible symmetry and role for Euclidean space-time in cosmology'',
 {\it New Astronomy} {\bf 89} (2021) 101635.
\end{thebibliography}
\end{document}